\documentclass[acmlarge]{acmart}

\usepackage{booktabs} 

\usepackage[ruled]{algorithm2e} 

\SetAlFnt{\small}
\SetAlCapFnt{\small}
\SetAlCapNameFnt{\small}
\SetAlCapHSkip{0pt}
\IncMargin{-\parindent}

\acmVolume{0}
\acmNumber{0}
\acmArticle{0}
\acmYear{2018}
\acmMonth{0}
\acmArticleSeq{0}

\setcopyright{usgovmixed}

\acmDOI{0000001.0000001}

\begin{document}
\title{Mobile Money: Understanding and Predicting its Adoption and Use in a Developing Economy}

\author{Simone Centellegher}
\affiliation{%
  \institution{Fondazione Bruno Kessler, University of Trento, Vodafone Research}
  \city{Trento} 
  \country{Italy} 
}
\email{centellegher@fbk.eu}

\author{Giovanna Miritello}
\affiliation{%
  \institution{Vodafone Research}
  \city{London} 
  \country{United Kingdom} 
}
\email{giovanna.miritello@vodafone.com}

\author{Daniel Villatoro}
\affiliation{%
  \institution{Vodafone Research}
  \city{London} 
  \country{United Kingdom} 
}
\email{daniel.villatoro@vodafone.com}

\author{Devyani Parameshwar}
\affiliation{%
  \institution{Vodafone Group}
  \city{London} 
  \country{United Kingdom} 
}
\email{devyani.parameshwar@vodafone.com}

\author{Bruno Lepri}
\affiliation{%
  \institution{Fondazione Bruno Kessler}
  \city{Trento} 
  \country{Italy} 
}
\email{lepri@fbk.eu}

\author{Nuria Oliver}
\authornote{This is the corresponding author}
\affiliation{%
  \institution{Vodafone Research}
  \city{London} 
  \country{United Kingdom} 
}
\email{nuria.oliver@vodafone.com}

\begin{abstract}
Access to financial institutions is difficult in developing economies and especially for the poor. However, the widespread adoption of mobile phones has enabled the development of mobile money systems that deliver financial services through the mobile phone network. Despite the success of mobile money, there is a lack of quantitative studies that unveil which factors contribute to the adoption and sustained usage of such services. In this paper, we describe the results of a quantitative study that analyzes data from the world{'}s leading mobile money service, M-Pesa. We analyzed millions of anonymized mobile phone communications and M-Pesa transactions in an African country. Our contributions are threefold: (1) we analyze the customers' usage of M-Pesa and report large-scale patterns of behavior; (2) we present the results of applying machine learning models to predict mobile money adoption (AUC=0.691), and mobile money spending (AUC=0.619) using multiple data sources: mobile phone data, M-Pesa agent information, the number of M-Pesa friends in the user's social network, and the characterization of the user's geographic location; (3) we discuss the most predictive features in both models and draw key implications for the design of mobile money services in a developing country. We find that the most predictive features are related to mobile phone activity, to the presence of M-Pesa users in a customer's ego-network and to mobility. We believe that our work will contribute to the understanding of the factors playing a role in the adoption and sustained usage of mobile money services in developing economies.
\end{abstract}

%
%


%
%




\maketitle

\renewcommand{\shortauthors}{S. Centellegher et al.}

\section{Introduction}
Nowadays, there are approximately 2 billion \textit{unbanked} individuals worldwide\footnote{http://www.worldbank.org/en/programs/globalfindex}. The \textit{unbanked} are defined as those adults who are not bank account holders or do not have access to a financial institution. Access to financial institutions is difficult in developing economies and especially for the poor, due to the low penetration of financial services in such countries, particularly in rural areas. 
For the last ten years, access to financial services by \textit{unbanked} individuals has been expanding partly because of the rapid growth in the adoption of mobile money services \cite{demirgucc2015global}. 
The widespread adoption of mobile phones, including in developing countries, has enabled the rise of mobile money services. Mobile money bridges the gap between the cash and digital economies, enabling those without access to banks to load cash in a mobile wallet and transact digitally using money transfers, deposits and withdrawals of money, bill payments, etc. through the mobile phone network. In developing countries, these services have been extensively successful, led by the example of the world's leading mobile money service M-Pesa launched in Kenya in 2007 and deployed today in 8 countries.

Mobile money services have had a tremendous positive impact on people's lives and have contributed to increased financial inclusion and economic growth, absorbing financial shocks and reducing poverty \cite{burgess2005rural,aportela1999effects,cull2014financial}. In recent work, Suri \emph{et al.} estimate that the mobile money service M-Pesa lifted 2\% of Kenyan households out of poverty \cite{suri2016long}.

At the end of 2014, the number of registered mobile money accounts was larger than the number of bank accounts in more than 15 developing countries \cite{scharwatt2014state}.
As of today, 277 mobile money services are active worldwide, 35 of which have over 1 million active accounts\footnote{State of the Industry Report on Mobile Money, Decade Edition: 2006 - 2016. GSMA. 2017.}.

Despite the success of mobile money services, a few challenges have hindered their deployment \cite{ifc2010m}, including the demographic and economic context of the country at hand, the development of effective marketing strategies, the deployment of a well-structured agent network and the existence of country regulations. Product adoption and customer engagement are key indicators of the adoption of mobile money services, such that the number of active accounts is used to understand how customers are adopting mobile money services \cite{scharwatt2014state}. Thus, it is somewhat surprising that there has been a lack of quantitative studies that shed light on the factors that play a role in the adoption and usage of mobile money services. 

In this paper, we aim to fill this gap and describe the results of a quantitative study that analyzes data from the world's leading mobile money service, M-Pesa. We report our findings after analyzing 140 million of mobile phone records and more than 27 million of mobile money (M-Pesa) transactions in an African country, over a period of 6 months. Through the analysis of this data, we focus on understanding the key drivers of M-Pesa adoption and intensity of usage. 

The main contributions of this paper are three-fold: 
(1) we first analyze large-scale patterns of M-Pesa usage --including the types of services used-- and compare communication and money transfer flows;
(2) we report the results of building two machine learning-based models to predict, three months into the future, mobile money adoption and mobile money spending. The models use multiple sources of data, including mobile phone data, M-Pesa agent information, the number of M-Pesa friends in the user's social network, and the type of geographic location (\textit{i.e.} rural vs urban) where the mobile activity takes place; finally, 
(3) we discuss the most predictive features in each of the models and draw implications for the design of mobile money services in a developing country.

\section{Related Work: Mobile Money}
An extensive portion of previous work in the area of mobile money focuses on how to successfully implement and deploy mobile money services, and how to define national regulations that encourage financial inclusion while minimizing financial risks and fraud. Such studies analyze the factors that drive a successful mobile money implementation \cite{camner2009makes} and how mobile money deployments can reach a critical mass of customers in order to scale the service to a successful deployment \cite{mas2011scaling}. Other works focus on how policy-makers and practitioners can create and enable a favorable environment for mobile money to flourish \cite{dermish2011branchless} and how mobile money can promote financial inclusion in the developing world \cite{must2010mobile,donovan2012mobile}. For example, the main factors behind mobile money adoption and the economic impact of mobile money are described in \cite{suri2017mobile}. Most of these works target a specific kind of audience, namely policymakers and regulators.

From an ethnographic perspective, there have been several studies aimed at understanding the adoption of mobile money services in low-educated and low-income individuals via field studies \cite{morawczynski2009exploring}, interviews and qualitative user studies \cite{medhi2009mobile,suri2017mobile}. 
Within the Human Computer Interaction (HCI) literature, several works have studied how customers interact with mobile money and the impact of their design on customer behavior \cite{medhi2009comparison,chiang2017understanding}.

Other works have analyzed mobile phone usage data --namely Call Detail Records (CDRs)-- to study individual and collective human financial behavior.  
Mobile phone data has been shown to be valuable to predict people's spending behavior \cite{singh2013predicting}, to infer their credit score \cite{san2015mobiscore} and the socio-economic status of a population  \cite{soto2011prediction,blumenstock2015predicting}. For example, in \cite{blumenstock2015predicting} the authors developed a model to map poverty in Rwanda, estimating the wealth of $1.5$ million customers from their mobile phone activity.

However, there are very few studies to date that analyze large-scale mobile money data and build quantitative statistical models of mobile money usage behavior. An industry report by CGAP \cite{cgap2013power} explores the key drivers of mobile money adoption by analyzing mobile phone usage data and the number of mobile money users in a customer's social network. Recently, in \cite{khan2016predictors} the authors developed predictive models of mobile money adoption from mobile phone usage data in three different developing economies. One of their key findings is that transfer learning does not perform well, and the results in the countries under study vary significantly: it seems that there is not a common set of behavioral features derived from mobile phone usage data, that consistently predict mobile money adoption in different countries. A preliminary follow-up work by the same authors aims to understand the differences in the key drivers of mobile money adoption between female/male and richer/poorer districts~\cite{khan2017determinants}.

Our work expands these previous works. We report the results of the analysis of millions of mobile money (M-Pesa) transactions in an African country. Moreover, we perform two machine learning-based tasks to predict mobile money adoption and usage from multiple sources, namely mobile phone data (CDRs), M-Pesa agent information, the number of M-Pesa friends in the user's social network and the type of geographic location (rural vs urban) where the activity takes place. We then discuss the most predictive features in each of the models and draw key implications for the design of mobile money services in a developing country. To the best of our knowledge, this is the first quantitative study done with data from M-Pesa, the worldwide leading mobile money service.

\section{M-Pesa, a mobile money transfer platform}
In 2007, the largest mobile operator in Kenya, Safaricom (owned partially by Vodafone) launched a new payment and money transfer service delivered through its mobile phone network, known as M-Pesa. M-Pesa stands for Mobile ``Pesa'', the Swahili word for money and it is the world's most successful mobile money service.

After a simple registration phase, requiring an official form of identification, the service allows its customers to perform a variety of services, including deposit money on their M-Pesa account associated with their mobile phone; transfer money via an SMS to another mobile phone user; withdraw cash from their M-Pesa account; purchase airtime and pay bills. To enable money deposits and withdrawals, M-Pesa runs and maintains an extensive agent network distributed on the territory. In fact, M-Pesa acts as a branch-less banking service where the ``ATMs'' are replaced by agents, which generally consist of already existing airtime resellers and retail outlets.

Given its successful launch and rapid growth in Kenya, other mobile phone providers launched their own mobile money services. Mobile money service solutions require the development of the actual mobile service, the availability of an agent distribution network, a solid marketing strategy to build consumer trust in the service, an investment in service integrity and the existence of well-defined supportive regulations \cite{hughes2007m}. 

As of today, M-Pesa has grown rapidly and the service is offered in 8 countries: the Democratic Republic of Congo, Egypt, Ghana, India, Kenya, Lesotho, Mozambique and Tanzania. 

In this paper, we report the results of the analysis of mobile phone and M-Pesa data from an African country where M-Pesa has the largest market share (42\%), in competition with three other mobile money providers. Mobile money is well adopted in this country which has an estimated poverty rate of 47\%, with about 12 million people still living in extreme poverty and many others living just above the poverty line. Taken together, these aspects make the country under study an interesting candidate to analyze mobile money services.

\section{Data}
In this work, we analyze pseudo-anonymized\footnote{GDPR Article 4(5)} mobile phone metadata (CDRs) and M-Pesa financial transactions from our country of study. The data is pseudo-anonymized, which means that all personal information is encrypted to preserve privacy. However, data for the same customer uses the same hashed ID, such that we can match the pseudo-anonymized IDs for the same customer over time and across different datasets. 

We use two datasets from two different time periods. The first dataset, $D_1$, contains pseudo-anonymized Call Detail Records (CDRs) of a random sample of 100,000 customers during a three month period from November 2016 to January 2017 ($T_1$). In total, dataset $D_1$ has more than 140 million CDR events. 

The second dataset, $D_2$, contains M-Pesa transactions, from April 2017 to June 2017 ($T_2$), for a total of more than 1.2 million randomly selected customers (including the customers in $D_1$) who generated approximately 27 million M-Pesa financial transactions. 

In the following, we describe both the CDRs and M-Pesa data that we analyze in this paper.

\subsection{CDRs dataset: voice and SMS}

The CDRs contain metadata about mobile phone calls (voice) and SMS events. For privacy reasons, no conversational or textual content is recorded. Any personal information available in the metadata (\textit{i.e.} actual phone number) had been previously pseudo-anonymized with an encrypted hash. 

The metadata available for each pseudo-anonymized phone call record is the following: \textit{callerID} (encrypted), \textit{calleeID} (encrypted), \textit{type of the call} (incoming/outgoing), \textit{timestamp} of when the call took place, its \textit{duration} and the \textit{cell tower ID} through which the phone call was routed. In the case of SMS, the following information is available: \textit{senderID} (encrypted), \textit{recipientID} (encrypted), \textit{SMS type} (outgoing messages), \textit{timestamp} and \textit{cell tower ID} of the cell tower that routed the SMS. Unfortunately, the mobile phone operator did not provide us with the incoming SMS. Note that the encryption is consistent for the same callerID and calleeID, such that we can match the data from the same customers in different datasets.

In addition, we were provided with the mapping between cell tower IDs and cell tower locations (latitude and longitude) together with the classification of the area in which each cell tower is located (urban, suburban or rural). 

\subsection{M-Pesa dataset}

The M-Pesa dataset contains information on mobile financial transactions which include the pseudo-anonymized \textit{senderID}, the pseudo-anonymized \textit{recipientID}, the \textit{type} of transaction, the \textit{amount} of money in the transaction, the \textit{timestamp} and the \textit{cell tower ID} of the cell tower that routed the SMS associated with the mobile money transaction.

In addition to the M-Pesa transactions, we had access to the agent network and the agent locations, identified by the nearest cell tower to the actual location of the agent (\textit{cell tower ID}).

Finally, we were provided with information on whether all the individuals making and receiving calls/SMS in dataset $D_1$ are registered M-Pesa users or not.

\section{Descriptive Data Analysis}\label{sec:desc_analysis}
In the literature, there is a large number of studies based on the results of analyzing CDR datasets in a variety of domains, including socio-economic status and credit score inference \cite{soto2011prediction,blumenstock2015predicting,frias2013forecasting,san2015mobiscore}, natural disasters \cite{lu2012predictability, pastor2014flooding}, public health \cite{wesolowski2012quantifying,wesolowski2015impact}, transportation \cite{ccolak2016understanding}, and crime \cite{bogomolov2014once}. Interestingly, there are very few studies that have quantitatively analyzed mobile money-based financial transactions \cite{economides2017mobile}. Therefore, here we report a detailed descriptive analysis, mainly focused on the M-Pesa financial transactions.

\mbox{ } \\ 
\textbf{M-Pesa transaction types. }
M-Pesa provides different kinds of financial services. The most commonly used services in our dataset ($D_2$) are the following: (i) \textit{Customer Transfer}: direct money transfer between two M-Pesa customers; (ii) \textit{Pay Utility}: utility billings; (iii) \textit{Deposit at Agent Till}: actual deposit of real money at an agent to top up the M-Pesa account; (iv) \textit{Customer Withdrawal at Agent Till}: withdraw of real money from an agent from an M-Pesa account; and (v) \textit{Customer Airtime Purchase}: mobile phone top up. These types of transactions represent more than 91\% of all transactions in $D_2$.

\begin{figure}
\includegraphics[width=0.7\linewidth]{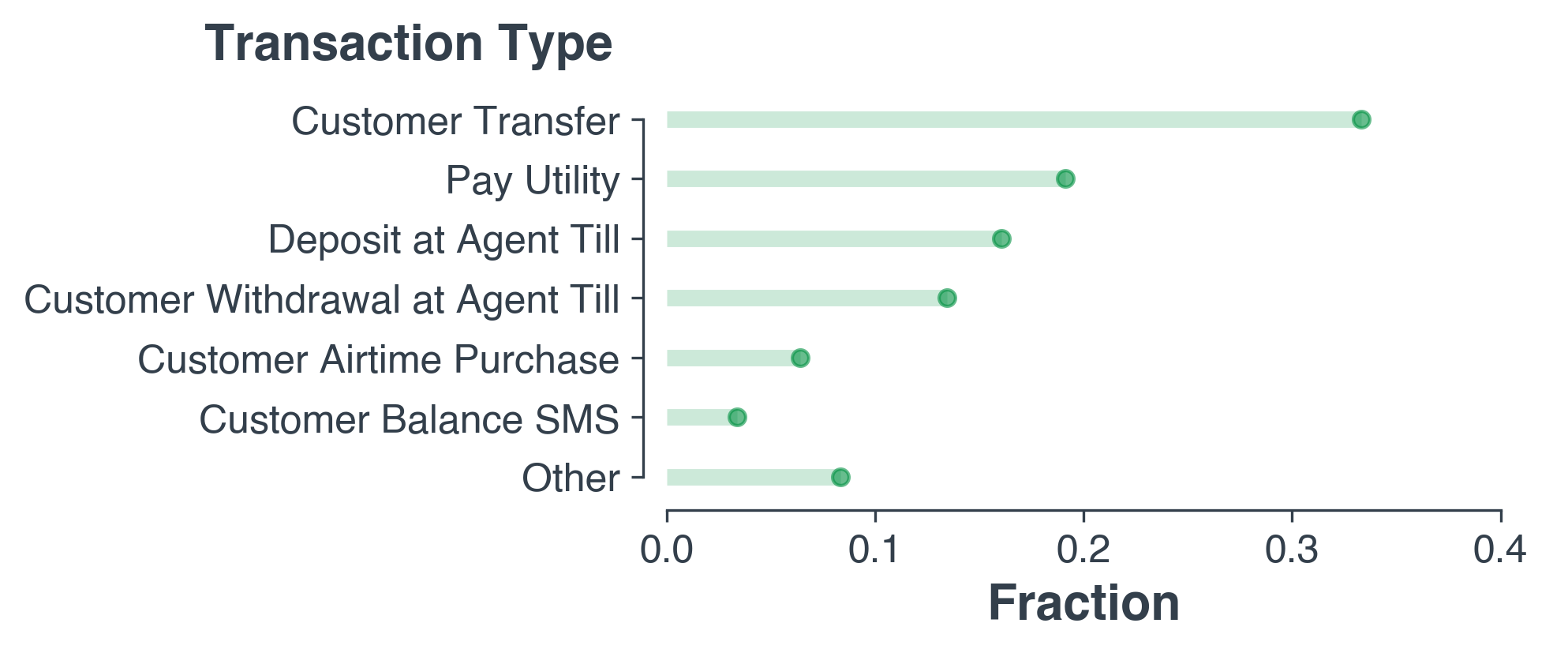}
\caption{Fraction of M-Pesa transactions type in dataset $D_2$}
\label{fig:trans_type}
\end{figure}

The most frequently performed transactions by our customers  are Customer Transfers, namely peer-to-peer (P2P) transactions and payments for utility billings, which is consistent with the original purpose of M-Pesa (money transfers). 

Figure~\ref{fig:trans_type} shows the percentages of transactions, divided by type, made by our customer base in $D_2$. Note that more than 30\% of all the transactions are P2P transactions (Customer Transfer). 

\begin{figure}
\centering
\begin{minipage}{.35\textwidth}
  \centering
  \includegraphics[scale=0.6]{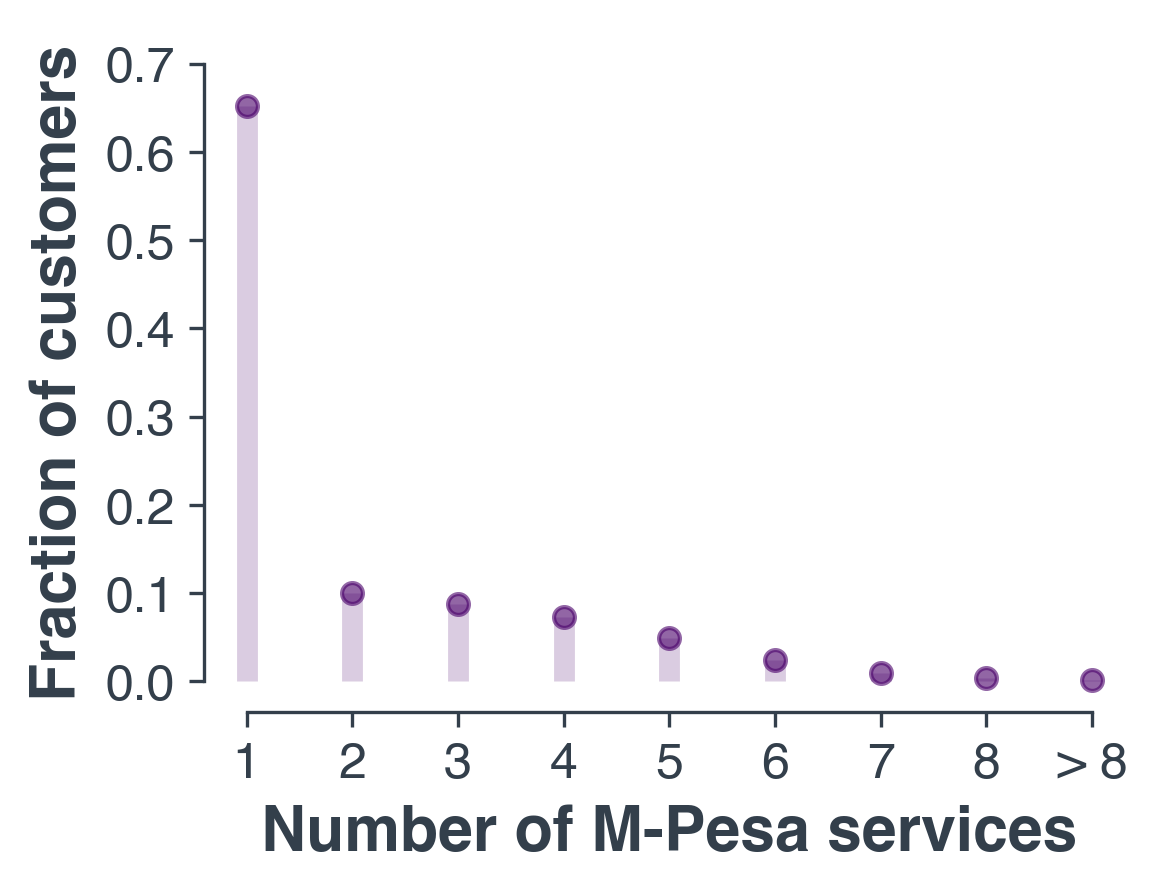}
  \captionof{figure}{Number of M-Pesa services used by customers in $D_2$}
  \label{fig:mpesa_channels}
\end{minipage}%
\hspace{0.5cm}
\begin{minipage}{.60\textwidth}
  \centering
  \includegraphics[width=0.7\linewidth]{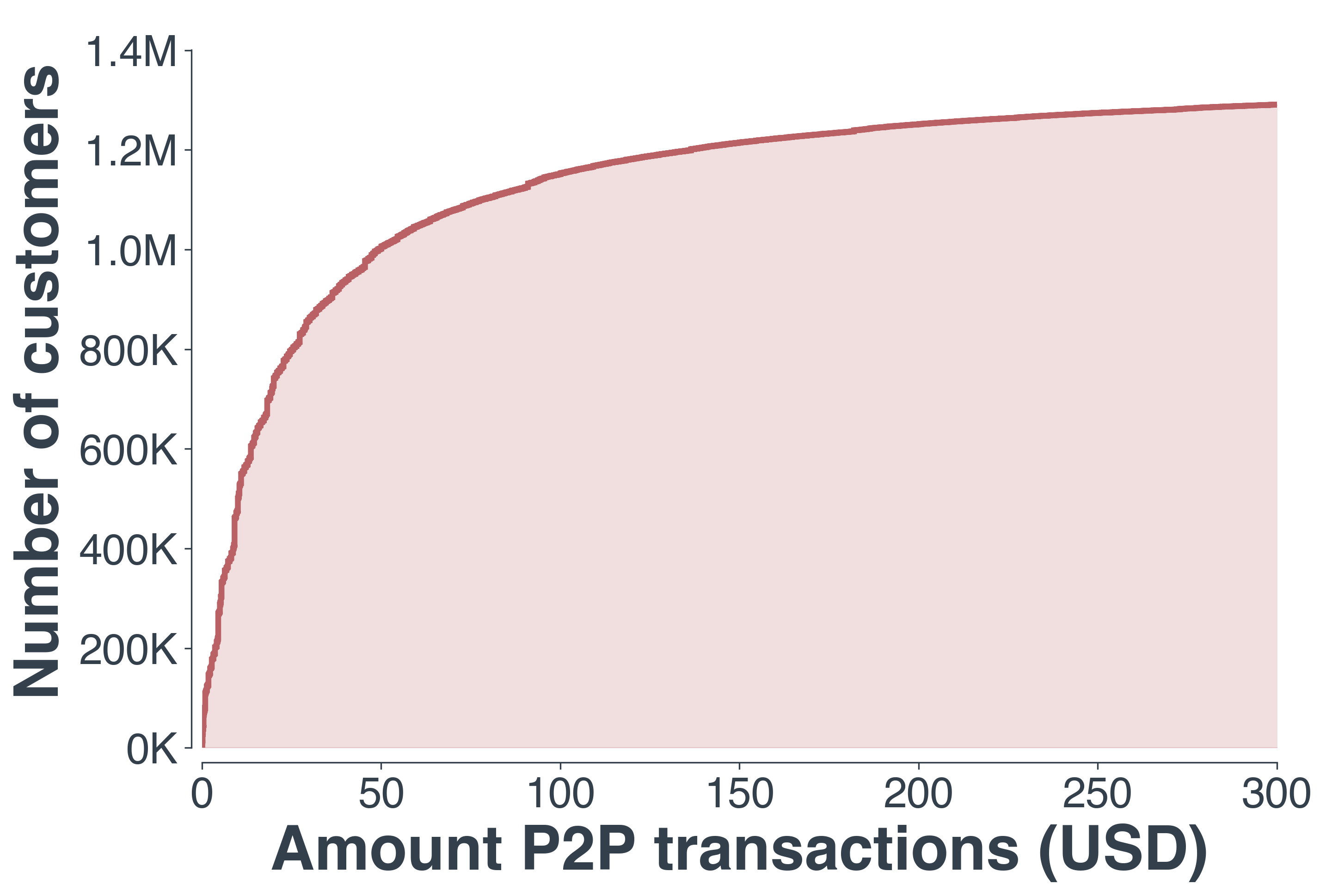}
  \captionof{figure}{CDF of the amount of money transferred via P2P transactions made by M-Pesa customers in dataset $D_2$ from April 2017 to June 2017 (truncated at 300 USD for visual purposes).}
  \label{fig:amount_user}
\end{minipage}
\end{figure}

\mbox{ } \\ 
\textbf{P2P transactions. }
Since P2P transactions represent a direct money transfer between M-Pesa customers, they can reveal interesting insights about customer behavior. Figure~\ref{fig:amount_user} shows the amount of money transferred via P2P transactions by M-Pesa customers in dataset $D_2$ from April 2017 to June 2017.
The monetary amount in P2P transactions is distributed as a long-tailed with the large majority of the customers transferring small amounts of money. 

While the absolute numbers may appear to be small, it is important to contextualize the data to the country under study. In fact, our country ranks in the lowest tier of the Human Development Index (HDI), a composite statistic used by the United Nations (UN) that takes into account the country life expectancy, education, and per capita income indicators.

\mbox{ } \\ 
\textbf{M-Pesa services. } 
Our next analysis focuses on the type of usage that customers make of M-Pesa services. As depicted in Figure~\ref{fig:mpesa_channels}, more than 65\% of all customers in $D_2$ use M-Pesa \textit{only for one purpose}. This percentage drops to 10\% for customers who use two M-Pesa services. 

In the group of customers who only use M-Pesa for one purpose, 89\% of their transactions correspond to either \textit{Customer Transfer} (P2P transfer) or \textit{Deposit at Agent Till}. This means that this group of customers is mostly using M-Pesa either for saving or for sending money. Therefore, we can consider these customers to act as ``sources'' of money. 

It is also worth mentioning that within this group, less than 1\% of the transactions are made to pay utility bills. In other words, a person is unlikely to be using M-Pesa if the only service they need is to pay utility bills. 

In the case of customer who use M-Pesa for two purposes (10\% of the customers), the most used services in this group of customers are \textit{Customer Withdrawal at Agent Till} (28.8\% of all transactions), followed by \textit{Pay Utility} (17.1\% of all transactions), \textit{Customer Transfer} (16.6\% of all transactions) and \textit{Deposit at Agent Till} (14.6\% of all transactions).

In fact, \textit{Customer Withdrawal at Agent Till} is the most common transaction and appears in $70\%$ of all transaction pairs, which implies that this subset of customers use M-Pesa to withdraw real money. Therefore, we can consider these customers to be ``recipients'' of money.

Given that poverty is particularly pervasive in the rural areas of the country of interest\footnote{``About 10 million people in the rural population live in poverty, and 3.4 million live in extreme poverty, compared 
to less than 1.9 million living in poverty and 750,000 people in extreme poverty in the urban areas in the country of study''. World Bank.}, we analyzed where the M-Pesa transactions originate looking at the classification of the area in which each cell tower is located (rural, urban, suburban). Interestingly, single-purpose M-Pesa users have statistically significantly more transactions that originate in urban areas ($0.447$) with respect the subset of customers that use M-Pesa for two purposes ($0.397$) with two-proportion z-test, $z=118.04$, $p<0.001$. The opposite is true for the transactions originated in rural areas, with the ``recipients'' of money ($0.47$), having more transactions in rural areas when compared to single-purpose users ($0.418$) with two-proportion z-test, $z=125.10$, $p<0.001$.

\mbox{ } \\ 
\textbf{Calls and money flows. }
Our next analysis focuses on where and how mobile phone interactions and M-Pesa transactions take place inside the country analyzing the CDRs data in $D_1$ and the M-Pesa transaction data in $D_2$.

We focused our analysis at the district level (169 districts). In order to infer the flows of mobile phone interactions and M-Pesa transactions, we first aggregated the transactions at the district level. Then, since we did not have the home location of the customers, we inferred it by detecting the most commonly used antennas during nighttime (7pm - 7am).

After this geolocation phase, we concentrated our analysis only on the M-Pesa P2P transactions of customers for whom we also have mobile phone interactions. This left us with 3,182 customers who made 46,478 mobile phone interactions and 320 M-Pesa P2P transactions between each other. Even with this reduced dataset, we observe an interesting difference between the geographic flows of phone calls and M-Pesa transactions: while the majority of calls take place \textit{intra-district} (63\%), the majority of M-Pesa P2P transactions take place \textit{inter-district} (72.5\%). This means that money is flowing at longer distances when compared to mobile phone calls.

\section{Predicting M-Pesa adoption and customer spending}
In this section, we describe the machine learning-based predictive models that we built to predict M-Pesa adoption and intensity of usage in the future, from features computed from mobile phone usage patterns, the M-Pesa agent network information, the number of M-Pesa friends in the user's communication graph, and characterization of user's geographic location. First we describe the features we extracted, followed by how we set up the tasks for predicting mobile money adoption and future customer spending. Then, we report the performance of the machine-learning models and describe the most predictive features in each task.  

\subsection{Feature extraction}\label{sec:feat}
We hand-crafted 77 features that were used to set up two machine learning tasks, namely (i) predicting M-Pesa usage and (ii) predicting M-Pesa expenditure three months in the future. The features we present here were designed to offer a comprehensive picture of customer behavior.

Since we did not have access to incoming SMS data, we computed all our metrics using voice CDRs. We grouped the computed features in 11 groups, according to the types of user behavior that they capture, as described below. 

\mbox{ } \\ 
\textbf{G1. Call interactions. }
The features in this group are based on the number of call interactions made by each customer, namely the total number of interactions (\textit{nr\_interactions}), the number of incoming interactions (\textit{in\_interactions}) and outgoing interactions (\textit{out\_interactions}). These features were computed over the entire three month period of $D_1$. We differentiated the features for weekends vs. weekdays in order to unveil possible differences in behavior during the working week when compared to the weekend. Finally, we computed the percentage of nocturnal calls (\textit{perc\_noct}), \textit{i.e.} the percentage of calls made between 7pm and 7am.

\mbox{ } \\ 
\textbf{G2. Call duration.}
We also computed features that characterize the duration of phone calls, namely the total duration of all phone calls (\textit{call\_duration}), the duration of incoming calls (\textit{in\_call\_duration}) and the duration of outgoing calls (\textit{out\_call\_duration}). Again, we computed these features as the total duration over the entire three month period $T_1$, differentiating between weekends/weekdays and day/night duration. 

\mbox{ } \\ 
\textbf{G3. Call regularity. }
In order to characterize the regularity of mobile phone usage over time, we computed the following features over the entire three month period $T_1$: the average (\textit{time\_int\_avg}), the median (\textit{time\_int\_median}) and the standard deviation (\textit{time\_int\_std}) of the inter-event time, \textit{i.e.} the time between two consecutive calls. 

\mbox{ } \\ 
\textbf{G4. Daily and weekly calls. }
Mobile phone usage is not homogeneously distributed with time \cite{aledavood2015daily}. Hence, we defined features to capture possible differences in phone usage behavior at a daily and a weekly levels. Specifically, we computed the average and the standard deviation of the number of daily calls (\textit{daily\_calls\_avg, daily\_calls\_std}). We then divided these metrics by weekdays and weekends (\textit{e.g.,} \textit{daily\_calls\_avg\_weekend}, etc). Moreover, we computed the daily average and standard deviation using the duration of the calls (\textit{daily\_call\_duration\_avg, daily\_call\_duration\_std}).
Finally, we computed the average and the standard deviation of the number and calls duration at weekly level (\textit{e.g.,} \textit{daily\_calls\_avg\_weekend}, \textit{weekly\_ calls\_avg\_weekend}, etc.).

\mbox{ } \\ 
\textbf{G5. Mobile phone usage. }
The last group of features related to mobile phone usage are the total number of active days (\textit{active\_days}) --which measures how active a customer is-- and the number of sent SMS\footnote{We only have access to outgoing SMS data.} (\textit{SMS\_sent})
 
\mbox{ } \\ 
\textbf{G6. Ego-network. }   
It is reasonable to speculate that the social network of an individual (ego) plays a role in his/her usage of M-Pesa.
The next three groups of features are designed to characterize network measures at the individual level. A simple measure widely used for quantifying the dimension of an individual's social network, or ego-network, is the network's degree. The ego-network features that we computed are: (i) the number of unique contacts (\textit{degree}), (ii) the number of unique individuals contacted by the ego (\textit{out\_degree}), and (iii) the number of unique individuals that contacted the ego (\textit{in\_degree}).

\mbox{ } \\ 
\textbf{G7. Social relationships. }
The features in this group are designed to measure additional characteristics of the social relationships as captured by the call graph. First, we defined a metric that characterizes the number of contacts that are responsible for 50\% of all the interactions for each individual (\textit{nr\_contacts\_50\_perc\_int}). This metric allows us to better characterize the customers' inner circle \cite{hill2003,dunbar2009social}, \textit{i.e.} their closest relationships. 
Second, we define a feature to measure the \textit{diversity} of contacts $D_{social}$ of each individual, given by the formula:
\begin{equation}
D_{social}(i)= -\frac{\sum_{j=1}^k p_{ij} \log(p_{ij})}{\log k}
\label{eq:dsocial}
\end{equation}
where $k$ is the number of contacts of user $i$, $ p_{ij} = \frac{V_{ij}}{\sum_{j=1}^k V_{ij}}$ and $V_{ij}$ is the volume of calls between user $i$ and user $j$.
It gives a measure of how an individual diversifies his/her calling volume given his/her communication patterns. Customers that contact a high number of individuals with respect to their communication patterns will have a high value of $D_{social}$; conversely, customers that always interact with the same few contacts will have a low value of $D_{social}$.

\mbox{ } \\ 
\textbf{G8. M-Pesa registered friends. }
Given the influence of our social network on our behavior \cite{hill2006, christakis2013}, it is reasonable to hypothesize that the larger the number of contacts who are registered M-Pesa customers, the larger the probability that customer $i$ will use M-Pesa \cite{cgap2013power}. 
In this group of features, we compute the number of M-Pesa friends (\textit{mpesa\_friends}) and the percentage of M-Pesa friends (\textit{perc\_mpesa\_friends}) present in a customer's call graph. 

\mbox{ } \\ 
\textbf{G9. Mobility.}
For every CDR event, we have the geographical position of the antenna where the event originated. Using the geo-localization of the antennas, we are able to approximately reproduce the mobility of the customers. We thus develop several mobility metrics looking at two levels of spatial granularity: at the antenna level and at the district level (169 districts). 
Features in this group include the number of unique visited locations at the antenna and district levels (\textit{nr\_unique\_antennas} and \textit{nr\_unique\_districts}); the \textit{diversity} of places $D_{places}$ visited by the customers, which aims at representing the diversity of a customer's mobility. Similarly to Eq.~\ref{eq:dsocial}, it is defined as an entropy-based measure computed with the following formula:
\begin{equation}
D_{places}(i)= -\frac{\sum_{j=1}^k p_{ij} \log(p_{ij})}{\log k}     
\end{equation}
where $k$ is the total number of places visited by customer $i$, $ p_{ij} = \frac{V_{ij}}{\sum_{j=1}^k V_{ij}}$ and $V_{ij}$ is the number of times customer $i$ visits place $j$.
It gives a measure of how an individual diversifies the visited places in their mobility patterns. Intuitively, customers that visit a high number of places with respect their mobility patterns will have a high value of $D_{places}$; on the contrary, customers that always visit the same few places will have a lower value of $D_{places}$. We measure the diversity of places at the antenna level (\textit{H\_antennas}) and at the district level (\textit{H\_districts}). 

Another well known mobility measure is the \textit{radius of gyration} \cite{gonzalez2010}, which measures the typical distance travelled by a customer. It is defined as:
\begin{equation}
r_g = \sqrt{\frac{1}{N} \sum_{i=1}^{N}(r_i - r_{cm})^2}    
\end{equation}
where N is the total number of visited locations, $r_i$ is a location position recorded as longitude and latitude and $r_{cm}$ is the center of mass of the trajectories, defined as $r_{cm} = \frac{1}{N} \sum_{i=1}^{N} r_i$.
We compute this feature only at the antenna level because this granularity allows us to have a better representation of a customer mobility.

To have a sense of the number of the most frequent locations visited by a customer, we define \textit{nr\_antennas\_70\_perc}, which characterizes how many locations (antennas) account for 70\% of the total locations visited by a customer.

Finally we compute the \textit{daily average}, \textit{standard deviation} and \textit{median} of the unique number of antennas and districts visited. We also differentiate these variables by weekends and weekdays, as we did with the features that characterize the communication patterns (groups G1 to G4). Moreover, we compute the weekly \textit{average} and \textit{standard deviation} of the unique number of antennas and districts visited by a customer. 
     
\mbox{ } \\ 
\textbf{G10. M-Pesa agent information. }
An agent represents the physical availability of the service and their presence in the territory enables the deposit and withdrawal of real money by M-Pesa customers. For each customer, we thus infer their most used antenna during the daily hours (8am - 6pm), which represents a proxy for their work location, and during the night hours (7pm - 7am), which represents a proxy for the home location. From these two locations, we compute the minimum distance from an agent (\textit{min\_dist\_agents\_km}) and the number of agents that are accessible within a 500m radius (\textit{num\_agents\_500m}), which give a sense of the availability of M-Pesa agents in a particular area.

\mbox{ } \\ 
\textbf{G11. Type of location. }
Since we have information about the type of location of each antenna, namely if the antenna is located in a rural, sub-urban or urban area, we define the \textit{home\_class} and the \textit{work\_class} features as follows: for each customer, we infer the most used antenna during the daily hours (8am - 6pm) and during night hours (7pm - 7am). We then classify the type of home and work location as rural, sub-urban or urban.

\subsection{Classification tasks}\label{sec:class}
The features previously described allow us to capture a fairly comprehensive picture of a customer's behavior which we use as input to two machine learning-based models to predict future M-Pesa adoption and intensity of usage.

Our two datasets $D_1$ and $D_2$ were collected in different time periods: the dataset $D_1$ from November 2016 to January 2017 ($T_1$), and the dataset $D_2$, three months later, from April 2017 to June 2017 ($T_2$).
Hence, we compute the 77 features using the data in $D_1$ and define our M-Pesa target variables using the data in $D_2$. With this setting, we are able to investigate whether and how \textit{past mobile phone behavior}, as captured by features from $D_1$, is related to the \textit{future usage of M-Pesa}, as captured by M-Pesa-based target variables defined in $D_2$. In the following sections, we describe the two classification tasks we developed, together with the training/testing set up and the evaluation method of our classification models.

\begin{table}
    
{
\centering
\begin{minipage}{0.47\textwidth}
   \centering
   \caption{Performance of different classifiers on Task 1, measured as Area Under the Curve (AUC), F1-score, precision and recall using 5-fold cross-validation}
    \begin{tabular}{@{}lrrrr@{}}
        \toprule
         & \textbf{AUC} & \textbf{F1} & \textbf{Precision} & \textbf{Recall} \\ \midrule
        \textbf{GBT} & 0.691 & 0.463 & 0.37 & 0.66 \\
        \textbf{SVM} & 0.690 & 0.462 & 0.34 & 0.74 \\
        \textbf{LR} & 0.688 & 0.460 & 0.37 & 0.61 \\
        \textbf{RF} & 0.687 & 0.461 & 0.37 & 0.61 \\ \bottomrule
    \end{tabular}
    \label{tab:model_sel1}
\end{minipage}
\hspace{0.5cm}
\begin{minipage}{0.47\textwidth}
   \centering
   \caption{Performance of different classifiers on Task 2, measured as Area Under the Curve (AUC), F1-score, precision and recall using 5-fold cross-validation}
    \begin{tabular}{@{}lrrrr@{}}
        \toprule
         & \textbf{AUC} & \textbf{F1} & \textbf{Precision} & \textbf{Recall} \\ \midrule
        \textbf{GBT} & 0.614 & 0.612 & 0.62 & 0.61 \\
        \textbf{SVM} & 0.619 & 0.613 & 0.62 & 0.62 \\
        \textbf{LR} & 0.610 & 0.608 & 0.61 & 0.61 \\
        \textbf{RF} & 0.608 & 0.603 & 0.61 & 0.61 \\ \bottomrule
    \end{tabular}
    \label{tab:model_sel2}
\end{minipage}
}
\end{table}

\mbox{ } \\ 
\textbf{Task 1: Predicting M-Pesa usage}\\
The first classification task focuses on predicting M-Pesa usage in the future, from past mobile phone usage. Hence, we would like to automatically identify users who are inactive in $T_1$ but will be active in $T_2$.

This task is valuable to the business in order to design campaigns for customer acquisition and determine how many customers to contact. Being able to automatically determine customers likely to use M-Pesa based on their mobile phone usage behavior is of great interest to the marketing teams. This allow them to design and target their customer acquisition and engagement campaigns appropriately. From a scientific perspective, it is interesting to understand whether and to which degree past mobile phone usage captures elements of human behavior that are predictive of future mobile money usage.

While all the customers in the dataset $D_1$ are registered M-Pesa users, only a fraction are \textit{active} customers. In fact of the 92,574 customers for whom we have CDR data, 70,269 are inactive M-Pesa customers and 22,305 are active M-Pesa users. We define a customer as \textit{active} if (s)he has carried out at least one M-Pesa transaction in dataset $D_2$.

We randomly split the set of customers in 80\% for the training set (74,059 customers) and 20\% for the test set (18,515 customers). The ratio between the active and inactive customers in the train/test sets is the same as in the original data. Note that the trained models only have access to the data from the customers in the training set. Hence, they are tested in completely unseen data. 

As a model selection task, we tested different classifiers including Random Forests (RFs) \cite{breiman2001}, Gradient Boosted Trees (GBTs) \cite{friedman2001}, Logistic Regression (LRs) and Support Vector Machines (SVMs) \cite{cristianini2000}. 
We report the average Area Under the Curve (AUC) value across test sets using 5-fold cross-validation for all classifiers. As seen in Table~\ref{tab:model_sel1}, the performance of the classifiers was comparable. We decided to use Gradient Boosted Trees (GBT) since they are flexible, easy to train and they allow us to assess the importance of the features in the model. Besides reporting the highest AUC score of $0.691$, the GBT model is trading some recall for higher precision, which means having more false negatives and less false positives. From a business perspective, we are interested in having a low number of false positives since we do not want to miss customers who will be inactive because the model labels them as active. This allows us to intervene with customer offers designed to help increase their engagement with M-Pesa.

Our model performs at par with the state-of-the-art \cite{khan2016predictors} with an AUC score of $0.691$. It is interesting to see that we have comparable results to previous work, even if we are predicting M-Pesa usage 3 months into the future.

\begin{table}[h!]
\caption{Sensitivity of the results of Task 1 to the window of time used to train the model using a Gradient Boosted Tree (GBT)}
\begin{tabular}{@{}crrrr@{}}
\toprule
\textbf{Window Size} & \textbf{AUC} & \textbf{F1} & \textbf{Precision} & \textbf{Recall} \\ \midrule
\textbf{3 months} & 0.691 & 0.463 & 0.37 & 0.66 \\
\textbf{2 months} & 0.681 & 0.463 & 0.36 & 0.63 \\
\textbf{1.5 months} & 0.682 & 0.467 & 0.35 & 0.69 \\
\textbf{1 month} & 0.666 & 0.464 & 0.35 & 0.69 \\
\textbf{2 weeks} & 0.665 & 0.468 & 0.34 & 0.75 \\ \bottomrule
\end{tabular}
\label{tab:window}
\end{table}

\mbox{ } \\ 
\textbf{Training Window}
Next, we analyze the sensitivity of the results to the size of the time window used to train the model. As shown in Table~\ref{tab:window}, we trained our Gradient Boosted Tree classifier on different time windows starting from the entire period of 3 months from November 2016 till January 2017 (01/11/2016 - 31/01/2017) and shortening the training window until a period of time of 2 weeks (15/01/2017- 31/01/2017). We set the minimum amount of time to 2 weeks as we have several features computed at a weekly or weekend vs weekday levels. Having less than 2 weeks would yield very noisy features. As expected, the shorter the training window, the worse the model's performance. However, the drop in performance measured by the AUC value is not as large as expected. Taking into account the class imbalance, as we add more training data, precision increases while recall decreases, which means that the classifier has less false positives at the expense of having more false negatives, which is a desired behavior from a business perspective.

\mbox{ } \\ 
\textbf{Task 2: Predicting customer spending}\\
In the second classification task, we investigated whether the customers' past mobile communication behavior is related to their future mobile money spending behavior.
Focusing on the customers for which we have M-Pesa transaction data in $D_2$, we computed the total amount of money spent in the three month period under evaluation. Then, we created two classes of low and high spending customers by selecting the customers falling in the 25th and the 75th percentiles respectively, as shown in \cite{luo2017inferring}.

This task was defined by the business as they are interested in segmenting customers by their expenditure to design different marketing campaigns for high vs low spenders.

In this task, after the customer selection process, we have 7,990 customers in total: 3,995 in the \textit{high spender} class and 3,995 in the \textit{low spender} class. We then randomly split this set of customers into a training set (80\%, N=6,392) and a test set (20\%, N=1,598), preserving the same ratio of values of the target variable as in the original data.

We tested different state-of-the-art classifiers, including Random Forests, Gradient Boosted Trees, Logistic Regression and SVMs. 

Before testing the classifiers and given the sample and feature sizes, we applied a feature selection step in order to reduce the dimensionality of the feature space, lowering the risk of overfitting. We applied the Recursive Feature Elimination with Cross-Validation\footnote{http://scikit-learn.org/} (RFECV) method \cite{guyon2002gene}. After this step, we are left with 35 features.

As shown in Table~\ref{tab:model_sel2}, we also obtained comparable performance in all classifiers for this task. In this case, we report the results of a Support Vector Machine with RBF kernel and standardized features (z-score), since it provided slightly better results. We obtain an AUC value of $0.619$, which is the average value across test sets from 5-fold cross-validation.

\mbox{ } \\ 
\textbf{Threshold analysis.}
As previously described, we chose to split the M-Pesa customers into \textit{low} and \textit{high} spenders using the 25th and 75th percentiles, following the splits used in previous work  \cite{luo2017inferring}. We discuss in this section the sensitivity of the results for Task 2 as we change the percentile thresholds. As shown in Table~\ref{tab:sensitivity}, we started the analysis with a median split (50 - 50). From a business point of view, this is not a meaningful split since we are partitioning customers into low and high spenders with no margin.  For example, with a median value of 70 USD, this can lead in assigning a user that spends 69.5 USD into the low spender class and a user that spends 70.5 USD into the high spender class. Looking at Table~\ref{tab:sensitivity}, note that the larger the separation between the two classes, the better the model performance. In the case of 5th and 95th percentiles (5 - 95) split, we reach an AUC score of 0.715. The drawback, in this case, is the small sample size: the larger the class separation, the smaller the sample of people that it is possible to target. In Task 2, we decided to use the 25th and 75th because it gives us a fair trade-off balance between sample size, business value and model performance. Depending on the business needs, different thresholds might be used.

\begin{table}
\caption{Task 2 percentiles threshold analysis}
\begin{tabular}{@{}crrrrr@{}}
\toprule
\textbf{Threshold} & \textbf{AUC} & \textbf{F1} & \textbf{Precision} & \textbf{Recall} & \textbf{Sample size} \\ \midrule
\textbf{50 - 50} & 0.576 & 0.569 & 0.58 & 0.58 & 15,979 \\
\textbf{40 - 60} & 0.583 & 0.576 & 0.59 & 0.58 & 12,781 \\
\textbf{25 - 75} & 0.619 & 0.613 & 0.62 & 0.62 & 7,990 \\
\textbf{10 - 90} & 0.679 & 0.676 & 0.68 & 0.68 & 3,099 \\
\textbf{5 - 95} & 0.715 & 0.714 & 0.72 & 0.71 & 1,501 \\ \bottomrule
\end{tabular}
\label{tab:sensitivity}
\end{table}

\subsection{Feature Analysis}
In this section, we analyze the feature importance of the two classifiers built to tackle the two previously described tasks.

\begin{figure}
\includegraphics[width=\linewidth]{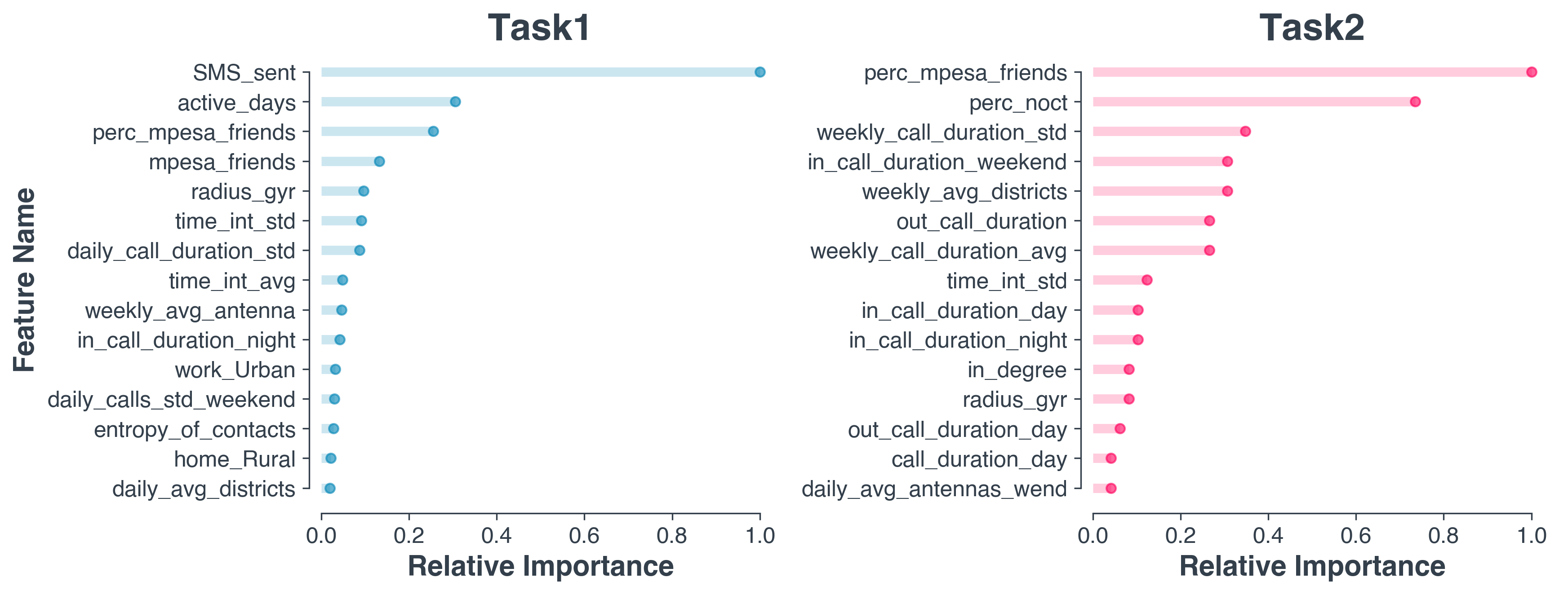}
\caption{Feature importance in Task 1: ``predicting M-Pesa usage" (left), and feature importance in Task 2: ``predicting M-Pesa expenditure" (right).}
\label{fig:imp_task}
\end{figure}

\mbox{ } \\ 
\textbf{Task 1. Predicting future M-Pesa usage.}
In Figure~\ref{fig:imp_task}, we show the top 15 most predictive features of the future M-Pesa usage. The feature importance is computed using the permutation importance method \cite{altmann2010permutation}.

The feature with the highest predictive power corresponds to the number of sent SMSs (\textit{SMS\_sent}), followed by the number of active days. Both of these features are proxies of intensity of mobile usage which seems to be an important factor in determining if a customer will use M-Pesa in the future. 

The next features in importance are related to the presence of M-Pesa users in a customer ego-network. As we can see, the \textit{perc\_mpesa\_friends} and \textit{mpesa\_friends} features impact the likelihood of the future M-Pesa usage of a customer. Thus, we observe that the customer's social network seems to affects the likelihood of using M-Pesa. In fact, social network influence has been already observed in other types of product adoption \cite{pan2011} and other types of human behavior \cite{christakis2013}.

Furthermore, mobility, and in particular the \textit{radius of gyration}, seems to play a role in determining future M-Pesa usage. 

While the rest of the features have lower impact, it is worth noting that we can find features related to the home and work location characterization together with other mobile phone usage features. Interestingly, we do not find any features related to the availability of M-Pesa agents. 

In order to further understand the association of the most important features and M-Pesa usage, we investigated the features' directionality. Thus, we trained a linear classifier (Logistic regression) and then analyzed the sign of the coefficient assigned to each of the top-15 features. As shown in Table~\ref{tab:feat_direction}, for this classification task, we have a positive association between features related to mobile phone usage and M-Pesa usage: for example, the larger the number of sent SMS (\textit{SMS\_sent}) and the larger the number of the days a customer is using his/her mobile phone (\textit{active\_days}), the higher the likelihood that a customer will use M-Pesa. As expected, the larger the number of M-Pesa users in a customer ego-network (\textit{mpesa\_friends}), the higher the likelihood that the customer will use M-Pesa in the future.

\begin{table}[]
\centering
\caption{Top-15 features coefficients and signs in our two classification tasks. The coefficients of the features are computed using a linear classifier (Logistic Regression). For the coefficient that are statistically significant we use the following notation: *$p<0.05$, **$p<0.01$, ***$p<0.001$.}
\begin{tabular}{@{}llll@{}}
\toprule
\textbf{Task 1} &  & \textbf{Task 2} &  \\ \midrule
\textbf{Feature} & \textbf{Coefficient} & \textbf{Feature} & \textbf{Coefficient} \\ \midrule
\textit{SMS\_sent} & +0.115*** & \textit{perc\_mpesa\_friends} & -0.278*** \\
\textit{active\_days} & +0.297*** & \textit{perc\_noct} & -0.248*** \\
\textit{perc\_mpesa\_friends} & +0.029** & \textit{weekly\_call\_duration\_std} & -0.033  \\
\textit{mpesa\_friends} & +0.250*** & \textit{in\_call\_duration\_weekend} & -0.331** \\
\textit{radius\_gyr} & +0.079*** & \textit{weekly\_avg\_district} & +0.220*** \\
\textit{time\_int\_std} & +0.178*** & \textit{out\_call\_duration} & +0.136 \\
\textit{daily\_call\_duration\_std} & +0.168*** & \textit{weekly\_call\_duration\_avg} & +0.226 \\
\textit{time\_int\_avg} & -0.119*** & \textit{time\_int\_std} & +0.010 \\
\textit{weekly\_avg\_antenna} & +0.112*** & \textit{in\_call\_duration\_day} & +0.153 \\
\textit{in\_call\_duration\_night} & -0.015  & \textit{in\_call\_duration\_night} & +0.134 \\
\textit{work\_urban} & -0.037*** & \textit{in\_degree} & +0.147** \\
\textit{daily\_calls\_std\_weekend} & -0.018 & \textit{radius\_gyr} & +0.104** \\
\textit{entropy\_of\_contacts} & +0.036*** & \textit{out\_call\_duration\_day} & -0.054  \\
\textit{home\_Rural} & +0.043*** & \textit{call\_duration\_day} & +0.026 \\
\textit{daily\_avg\_districts} & -0.001 & \textit{daily\_avg\_antennas\_wend} & -0.095* \\
\bottomrule
\end{tabular}
\label{tab:feat_direction}
\end{table}

\mbox{ } \\ 
\textbf{Task 2. Predicting customer spending.}
Figure~\ref{fig:imp_task} depicts the top 15 most predictive features of this classification task computed using the permutation importance method \cite{altmann2010permutation}.

Mobile activity features are again the most predictive for this task: three of the top 4 most predictive features are related to the intensity of usage of the mobile phone service. The feature with the highest predicting power regarding mobile activity is the percentage of nocturnal calls (\textit{perc\_noct}), followed by two features related to call duration (\textit{weekly\_call\_duration\_std} and \textit{in\_call\_duration\_weekend}). 

However, the most important feature is given by the percentage of M-Pesa users in the customer's social network (\textit{perc\_mpesa\_friends}). Other features related to the user's social network also play an important role, as expressed by the \textit{in\_degree}.

Finally, the impact of mobility is not negligible as the \textit{radius of gyration} and the average of the unique visited district in a week (\textit{weekly\_avg\_districts}) are among the most important features.

Again, to better understand the association of the most important features and M-Pesa spending, we investigated the features' directionality. We trained a Logistic Regression classifier, and then analyzed the sign of the coefficients assigned to each of the top-15 features. As shown in Table~\ref{tab:feat_direction}, the larger the mobility (given by the \textit{weekly\_avg\_district} and \textit{radius\_gyr}) of a customer, the higher the likelihood that the customer will be a high spender on M-Pesa. Previous work has also found a positive association between mobility and socio-economic status \cite{frias2012relation}. Presumably, high spenders on M-Pesa have higher socio-economic status than low spenders.

We also observe a positive correlation between \textit{in\_degree} and M-Pesa spending: the higher the \textit{in\_degree} of a customer, the higher the probability that the customer will be a high spender. Centrality measures in social networks, such as the degree of an individual, have been found in the literature to be correlated with his/her financial status \cite{luo2017inferring}.

Furthermore, we found that the larger the duration of incoming phone calls in the weekends, given by the  \textit{in\_call\_duration\_weekend} variable, the lower the probability of being a high spender. Previous work has reported that the lower the ratio between outgoing vs incoming calls, the lower the socio-economic status \cite{blumenstock2010mobile}.

It is interesting and somewhat counter-intuitive that the higher the percentage of M-Pesa friends in a customer's ego-network, the lower the probability of being a high spender. It could be possible that the higher the number of M-Pesa friends in the ego-network, the lower their socio-economic status of the person and hence the lower their spending on M-Pesa. We plan to investigate this hypothesis in future work.

\section{Discussion and Implications}

In this work, we have investigated whether and how \textit{past mobile phone behavior} is related to the \textit{future usage of M-Pesa}. We have shown that past mobile phone behavior is useful to predict M-Pesa usage 3-months into the future. We have also shown that is is possible to differentiate between high vs low-spenders in M-Pesa through the analysis of past mobile phone usage. Both predicting future usage of M-Pesa and determining the extremes (high vs low) in M-Pesa spending are meaningful tasks to carry out from a business perspective, as they enable the marketing teams to design different actions and campaigns that would be relevant to the customers depending on their levels of engagement with the mobile money platform.

In addition, we have carried out a feature importance analysis to shed light on the factors that play a role in the adoption and sustained usage of mobile money. 
We have found that intensity of mobile phone activity, characteristics of the customer's social network and to a smaller degree, mobility, have an impact in predicting M-Pesa usage.

Finally, we have provided two sensitivity analyses: one for the first task regarding the size of the training window and the other one for the second task in order to understand how the performance of our classifier varies depending on the percentile threshold used to select high vs low M-Pesa spenders. Depending on the business needs, different thresholds might be used.

From our qualitative and quantitative analyses, we draw several implications for the design of mobile money services.

\mbox{ } \\
\textbf{1. Money transfer prevails.} We found that the most frequently performed transactions are related to the original purpose of M-Pesa, namely money transfers. According to this finding, mobile money providers should prioritize the money transfer service over other services.

\mbox{ } \\
\textbf{2. Single purpose prevails.} More than 65\% of all customers in our dataset use M-Pesa only for one purpose. We identified ``sources'' of money --people who mainly use M-Pesa to deposit or to send money-- and ``recipients'' of money --people who use M-Pesa for two purposes, one of which is likely to be to withdraw money. The first group of customers --whose transactions often come from urban areas-- may have a higher money availability with respect to the second group --whose transactions generally originate in rural areas. Moreover, we found that a person is unlikely to be using M-Pesa if the only service (s)he needs is to pay utility bills. 

\mbox{ } \\
\textbf{3. Money travels farther than calls.} We found an interesting difference between the geographic flows of phone calls and M-Pesa transactions, with money flowing over longer distances with respect to mobile phone calls, which are a proxy for social interactions.

These three findings are important to take into consideration as mobile money systems expand their repertoire of services, because money transfer capabilities across long distances seem to be the key unmet need that these services fulfill. In addition, our analyses support previous work on how mobile money contributes to the economic development by enabling the transfer of money from urban (richer) to rural (poorer) regions \cite{morawczynski2009exploring}.     

\mbox{ } \\
\textbf{4. Past mobile phone behavior predicts mobile money adoption and expenditure}. Our multi-source mobile money adoption predictive model performs at par with the state-of-the-art \cite{khan2016predictors} (AUC=$0.691$, 38\% better than the baseline). It is interesting to see that we have comparable results to previous work even if we are predicting M-Pesa usage 3 months into the future. Our mobile money expenditure prediction model performs 32\% better than the baseline with an AUC of $0.619$. We cannot compare this result with other works since there is none in the literature.

The most predictive features are related to mobile phone activity (\textit{e.g.} \textit{SMS\_sent}, \textit{active\_days}) and to the presence of M-Pesa users in a customer's ego-network. Thus, we observe a relevant level of social virality in the likelihood of using M-Pesa, as also reported in another work on mobile money \cite{cgap2013power}. Moreover, the mobility of a customer is a non-negligible predictor as shown by the \textit{radius of gyration}.

\mbox{ } \\
\textbf{5. The value of the agent network}. Surprisingly, in the top 15 most predictive features of our models, we did not find features related to the distance and density of agents. Having access to an agent is necessary to be able to perform many of the offered mobile money financial services, including withdrawals and deposits of money (more than 30\% of all transactions in our datasets). There are multiple reasons that could explain our finding, including that the hand-crafted features do not capture the relevant aspects that the agent network plays. This can happen for example because of the approximation of the customers and the agents' location at the antenna level. However, it is also possible that there is already a sufficiently dense agent network covering the territory, such that the agent network does not play an important a role. 

\mbox{ } \\
Our findings and models also have business value as they enable mobile money service providers to better identify potential new customers of their services, anticipate consumption and understand the key drivers for mobile money adoption and usage. 
Previous research \cite{khan2016predictors} has found that CDR-based models developed in one country do not necessarily work in other countries. Hence, we are cautious to make any claims about the generalization capabilities of our findings to other countries. Our tasks and part of the features we devise are slightly different from the the aforementioned work, therefore we cannot directly compare the results. We leave to future work the investigation of how well our results generalize to different user samples, time periods and different countries.

\section{Conclusion and Future Work}
In this paper, we described the results of a quantitative study that analyzes data from the world's leading mobile money service, M-Pesa, together with data from a leading mobile phone operator in an African country. We analyze millions of anonymized mobile phone communications and M-Pesa transactions. 

We report an aggregate analysis of the customers' usage of M-Pesa and describe large-scale patterns of behavior. We also build two machine learning-based models to predict M-Pesa adoption and intensity of usage from a variety of data sources, including mobile phone usage, agent network information, the number of M-Pesa friends in the user's social network and the characterization of user's geographic region.

While our study is not exempt of limitations --including the fact that we have a random sample of customers and a 3 month gap between the two datasets-- we believe that this study is a first of its kind and will contribute to our understanding of the factors that play a role in the adoption and sustained usage of mobile money services in developing economies.

In future work, we plan to further study the impact of the agent distribution network, to replicate the study in a different country to assess the generalizability of our findings, to deploy and test the models in the real world, and to further investigate the impact of the social network (both call and M-Pesa network) on M-Pesa adoption and usage.

\section*{Acknowledgments}
We would like to thank the M-Pesa and Vodacom teams for their continued support and enthusiasm for this research.

\bibliographystyle{ACM-Reference-Format}
\bibliography{sample-bibliography}

\end{document}